\documentclass[graybox]{svmult}

\usepackage{mathptmx}       
\usepackage{helvet}         
\usepackage{courier}        
\usepackage{type1cm}        
\usepackage{makeidx}         
\usepackage{graphicx}        
\usepackage{multicol}        
\usepackage[bottom]{footmisc}

\makeindex             

\begin{document}

\title*{Synchrotron-radiation studies of topological insulators}
\author{Philip Hofmann}
\institute{Philip Hofmann \at Department of Physics and Astronomy, Aarhus University, Ny Munkegade 120, DK-8000 Aarhus~C, \email{philip@phys.au.dk}}%
%
\maketitle


\section{Introduction}
Topological insulators (TIs) are a recently discovered  \cite{Kane:2005b,Kane:2005c,Murakami:2006,Bernevig:2006b,Bernevig:2006,Fu:2007b,Fu:2007c,Moore:2007,Konig:2007,Hsieh:2008,Hsieh:2009,Zhang:2009} class of quantum materials that are currently attracting a lot of attention. The most interesting aspect of the TIs is, in fact, not that they are bulk insulators but that their surfaces support localized metallic states with some special properties, among others a characteristic spin texture. Most importantly, the existence of these metallic states is not a a surface property, it is required by the topology of the bulk band structure. While the details of the surface  (structure, reconstructions) still matter for the dispersion and Fermi contour of the surface states, their very existence is, in a sense, a bulk property.

Synchrotron-radiation based angle-resolved photoemission spectroscopy (ARPES) has been one of the most important tools to prove the existence of the topological surface states and spin-resolved ARPES has been able to confirm the predicted non-degeneracy of the states with respect to spin and the spin texture \cite{Hsieh:2008,Hsieh:2009,Xia:2009,Hsieh:2009c,Chen:2009,Hsieh:2009b}. The power of ARPES derives from the possibility of a direct spectroscopic view of the state's dispersion and of many-body effects such as the lifetime of the states. ARPES also stands out as an important technique because it has so far not been easy to probe the transport properties of the surface states directly \cite{Checkelsky:2009,Qu:2010,Analytis:2010,Analytis:2010b,Steinberg:2010,Checkelsky:2011,Chen:2011,Cho:2011,Xiu:2011,Butch:2010,Kim:2011a,Taskin:2012}. 
Superficially, this may appear surprising because a metallic surface on an insulating bulk crystal should be easy to detect. Unfortunately, this is not the case and many transport measurements have been dominated by the bulk states. The reason for this is the high bulk conductivity of the materials, which have a small band gap, many charged impurities and typically a high dielectric constant leading to poor screening between the impurities. The surface conductance, on the other hand, is not well known and lifetime ARPES studies could help to elucidate the processes that limit the lifetime of excited surface state carriers in transport.

This Chapter is not meant as a comprehensive review of topological insulators or of their study with ARPES but merely as an accessible introduction to the field for the non-specialist. We do not attempt to describe the historical development of the important discoveries in detail, nor do we show the original figures from such work. Instead, the emphasis is on a didactic presentation. We also discuss only one single material as an example for the important concepts, the widely-studied TI Bi$_2$Se$_3$ \cite{Xia:2009,Zhang:2009,Cheng:2010}. There are a number of excellent reviews on TIs already published \cite{Zhang:2008,Moore:2010,Hasan:2010,Kong:2011a,Qi:2011} and the reader is referred to these for a comprehensive description of the field.

In the following sections, we will first discuss the basic physical ideas that lead to the existence of topologically protected surface states on an insulator. Different explanations will given, ranging from a  simple hand-waving explanation to a practical description of how to determine if a material is a topological insulator. Following this, we will discuss how different TI states (bulk and surface) can be detected by ARPES and what their spectroscopic signature is. The chapter concludes with a discussion of how the electron-phonon coupling affects the lifetime of the surface states on Bi$_2$Te$_3$.

\section{Basic principles behind topological insulators}
\label{sec:1}

One of the most interesting questions in TI physics is how it is possible that the bulk electronic structure implies the existence of metallic surface states on a insulating material. In the following, we will encounter different pictures explaining this. Note that the term ``topological insulator'' might be slightly misleading because the materials are, in fact, not insulators with a large gap energy $E_g$ (in the sense that $E_g \gg k_B T$ at room temperature) but small gap semiconductors with $E_g \approx$300~meV. 

Metallic surface states on semiconductors are not an unusual phenomenon and simple electron counting arguments often dictate the surface of a truncated bulk semiconductor to be metallic \cite{Luth:1992}. On the other hand, the surfaces often undergo geometric reconstructions that change the periodicity parallel to the surface, leading to larger unit cells and non-metallic states. A simple model for such rearrangements is a Peierls distortion in a one-dimensional metallic chain that leads to an energy gain via a metal-insulator transition. On real surfaces, the picture is much more complex but the essence is that metallic semiconductor surfaces are usually just a coincidence \cite{Tosatti:1995}. They are not stable against, e.g. a structural rearrangement of the atoms.

\begin{figure}[b]
\sidecaption
\includegraphics[scale=1.0]{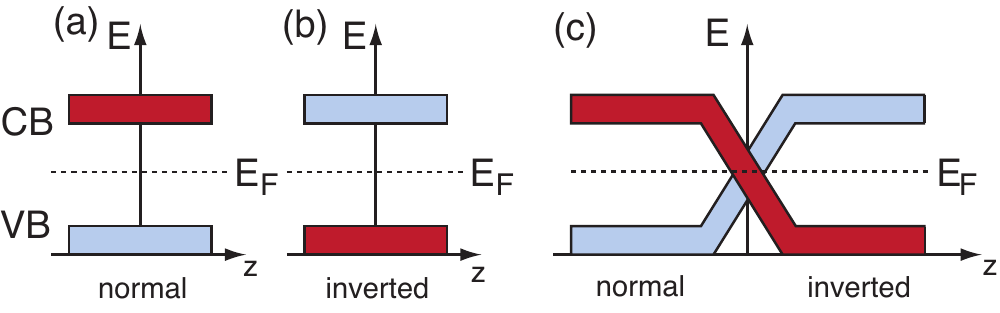}
%
%
\caption{Strongly simplified band diagram for semiconductors, showing the valence band (VB) and conduction band (CB) energies as a function of position (i.e. the edges signify the macroscopic ends of the sample). (a) 
Colour (parity) ordering of VB and CB in a normal semiconductor. (b) Semiconductor with an inverted band gap. (c) Joining a semiconductor with a normal band gap and one with an inverted band gap while maintaining the symmetry of the states gives rise to metallic interface states. }
\label{fig:1}       
\end{figure}

The existence of topological surface states, on the other hand, cannot depend on the structural details of the surface. A simple explanation for the stable existence of metallic states between two different semiconducting materials is given in Fig. \ref{fig:1}. Consider a ``normal'' semiconductor with a well-separated valence band (VB) and conduction band (CB) and the chemical potential (or Fermi energy $E_F$) in between. The VB and CB shall each have a characteristic colour, which for a ``normal'' material shall be 
blue and red, respectively. We will later discuss the meaning of the colour in more detail and we will see that it is related to the parity of the bands (we limit the discussion to materials with inversion symmetry).
Now suppose that we have another type of semiconductor, a TI, also with a well-defined band gap but an inverted order of the colours,
 i.e. a red VB and blue CB. If we join these two materials under the condition that we can only join states of identical character, the 
 blue and red states have to cross the Fermi energy at the interface, giving rise to two metallic states located there. This is the essence of a topologically protected metallic interface or surface state.

\begin{figure}[b]
\sidecaption
\includegraphics[scale=1.3]{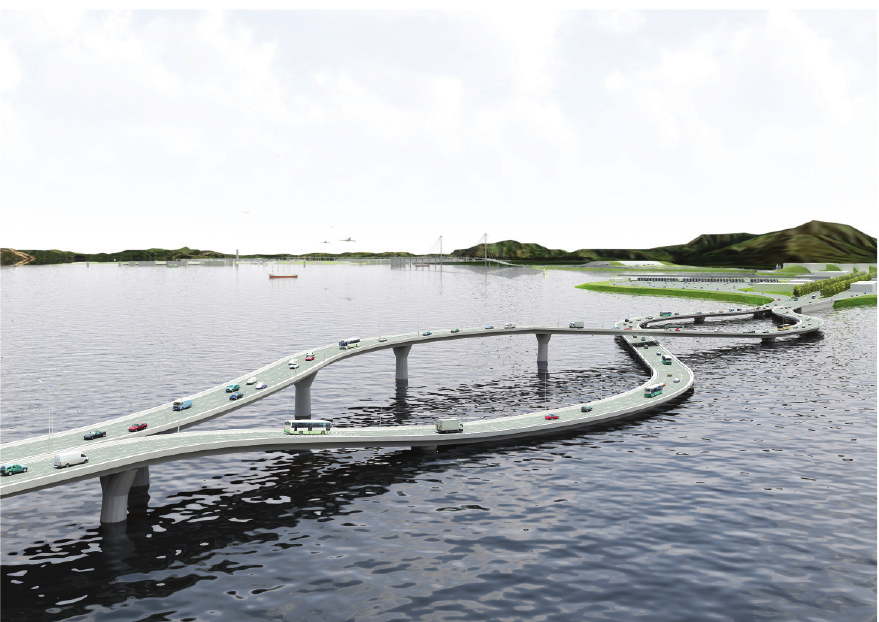}
%
%
\caption{Proposed ``flipper bridge'' between Hong Kong (left-hand traffic) and mainland China (right-hand traffic). Design by nl architects (www.nlarchitects.nl). Picture used with permission. }
\label{fig:2}       
\end{figure}

The crucial point is that the existence of the crossing is not a surface property but a bulk property. An excellent analogy is the situation that arises when two countries are to be joined by a road bridge, with the difficulty that the driving rules in one country enforce right-hand traffic and in the other left-hand traffic. A possible solution to this problem is the traffic flipper bridge shown in Fig. \ref{fig:2}, a proposal for a boundary crossing between Hong Kong with mainland China. In this case, the colour of the bands in Fig. \ref{fig:1} is representing the traffic side (left-hand vs. right-hand) and this is a bulk property of the two countries, requiring a node in the bridge. The solution is not unique: the node could be included in a more complex structure road layout, or there could be a higher number of crossings, as long as the total number of crossings is odd. But the bulk topology of the countries dictates the existence of at least one node somewhere near the border and the flipper bridge is the simplest topological solution to the problem.

\begin{figure}[b]
\sidecaption
\includegraphics[scale=.4]{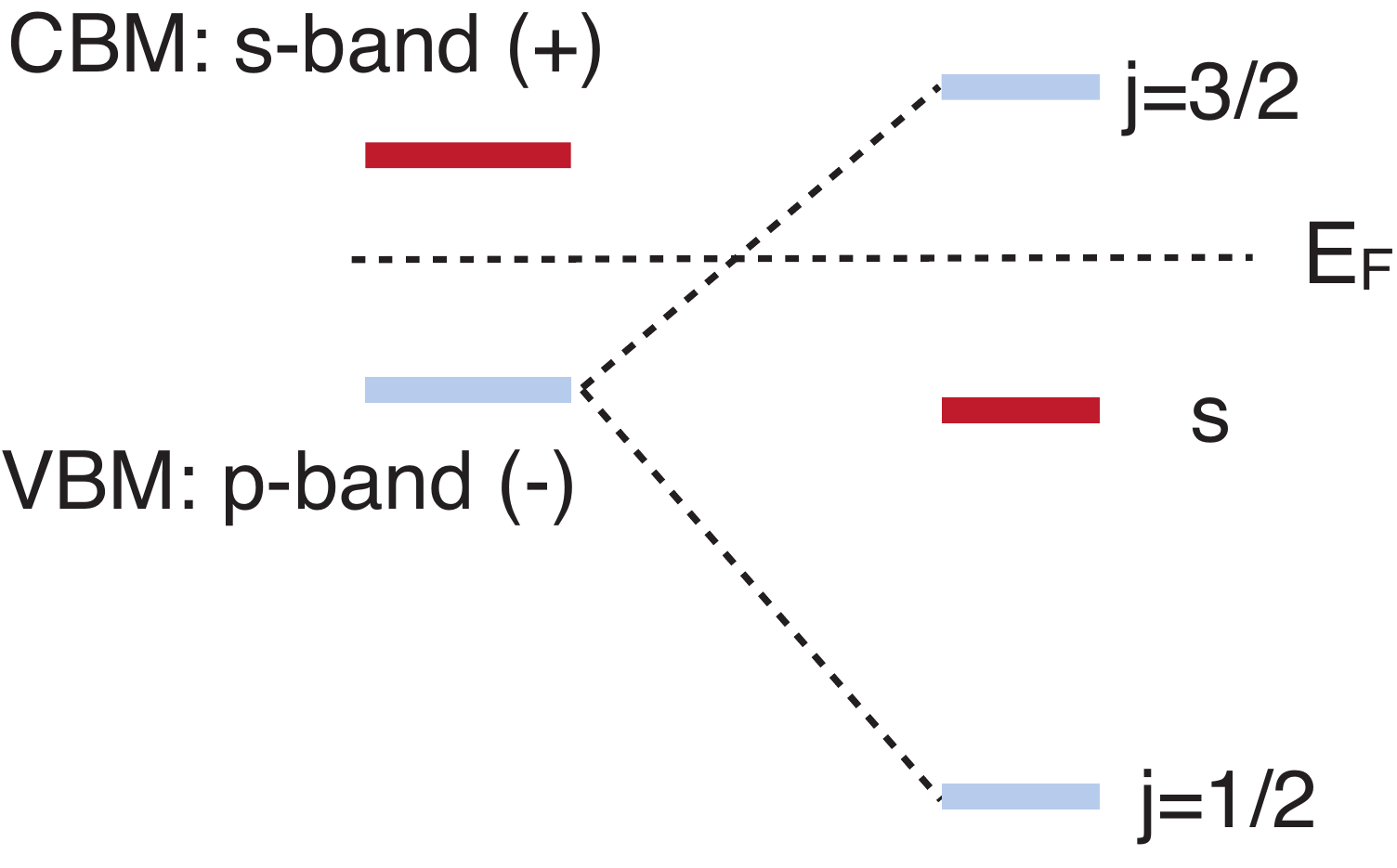}
\caption{Strongly simplified illustration of the effect of spin-orbit coupling on the symmetry of the valence band maximum (VBM) and conduction band minimum (CBM). For weak coupling, the VBM has $p$-orbital character and negative parity, the CBM has $s$ character and positive parity. Strong coupling can move the $j=3/2$ level above the $s$ level, leading to a parity-inverted band gap.}
\label{fig:2a}       
\end{figure}

In a material, the 
colour of the bands in Fig. \ref{fig:1} represents the parity of the bulk bands. In a TI, the bulk band structure has an inverted parity ordering due to the strong spin-orbit interaction.  A very simple illustration of this is given in Fig. \ref{fig:2a}. Consider the most important orbital angular momentum contribution to the electronic states of a semiconductor near the valence band maximum (VBM) and conduction band minimum (CBM), assuming a direct band at the Brillouin zone centre. In a ``normal'' material the VBM has mostly $p$ character (negative parity) whereas the CBM has mostly $s$ character (positive parity). If we now consider materials with a strong spin-orbit interaction, this lifts the degeneracy of the $p$ level, leading to the creation of  $j=1/2$ and a $j=3/2$ states. If the splitting is sufficiently strong, the $j=3/2$ state can move above the $s$ state, giving rise to the desired band parity inversion and the creation of a TI material.

These simple pictures give an intuitive explanation for the topological stability of interface states, but they have a number of severe shortcomings, as far as the details are concerned, and leave several questions open.  Fig. \ref{fig:2a}, for instance, explains the origin of the parity inversion but it is merely a picture derived from atomic states, not bands. This leads to some apparent contradictions, for example that the number of filled states is not conserved. Also, the pictures above may explain the existence of topologically protected interface states between two materials, but not the existence of such states at a surface, between a TI and vacuum. We come back to these points further down in the text.

\begin{figure}[b]
\includegraphics[scale=0.8]{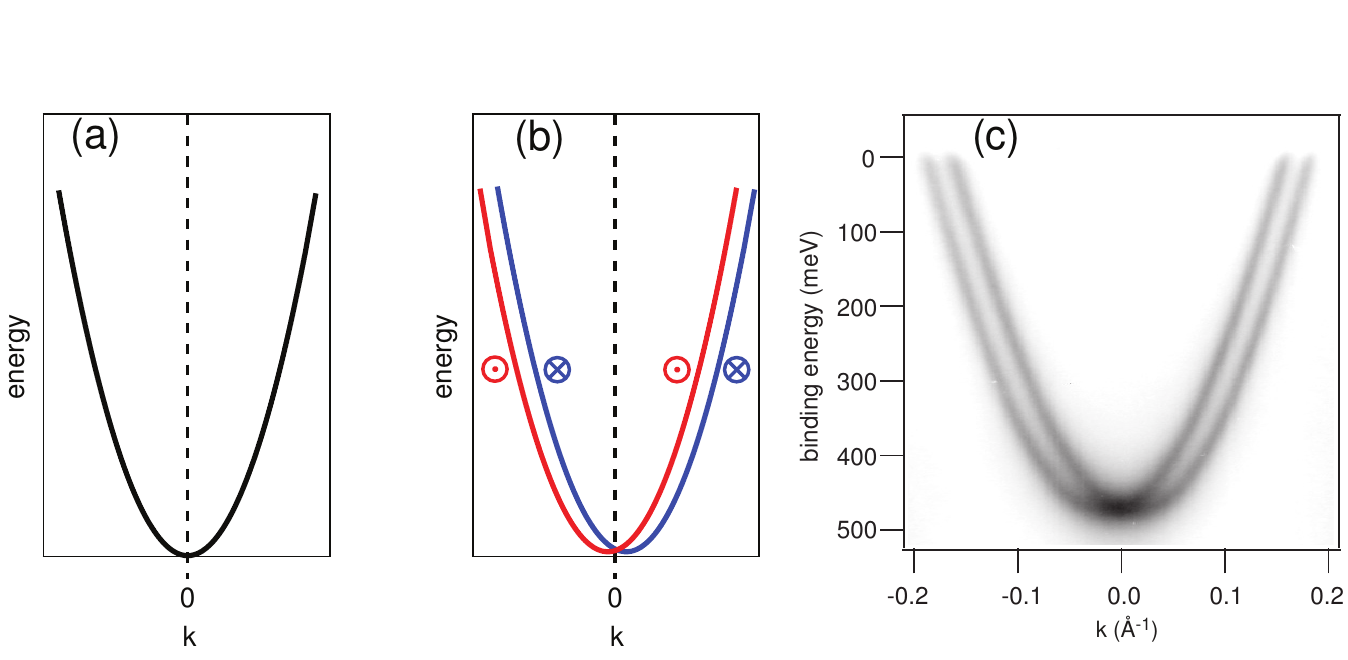}
\caption{(a) One dimensional cut through the two-dimensional dispersion of a free electron-like surface state, as found on the (111) surfaces of the noble metals. (b) The same state under the influence of strong spin-orbit coupling (Rashba effect). The spin-degeneracy is lifted and the spin directions of the individual branches are indicated by the arrow heads. (c) ARPES measurement of the spin-orbit split surface state on the Au(111) surface. High photoemission intensity is dark. After Ref.   \cite{Nechaev:2009}.}
\label{fig:3}       
\end{figure}

For now, we approach the idea of topologically protected surface states from another angle, by starting out with ``normal'' surface states in the presence of strong spin-orbit coupling. Consider a free electron-like surface state with a parabolic dispersion around the centre of the surface Brillouin zone (SBZ) $\bar{\Gamma}$, as shown in Fig. \ref{fig:3}(a). Such surface states are commonly found on the (111) surfaces of the noble metals Cu, Ag and Au \cite{Reinert:2001}. On metals,  such surface-localized solutions of the Schr\"odinger equation, can only exist in projected gaps of the bulk band structure. For the noble metal (111) surfaces such gaps are present around the Fermi energy at the $\bar{\Gamma}$ point because of the characteristic shape of the bulk Fermi surfaces which are almost spherical but have ``necks'' into the neighbouring Brillouin zone around the $L$ symmetry point \cite{Ashcroft:1976}. The projected bulk states and the gap are not shown in Fig. \ref{fig:3}(a).

A strong spin-orbit interaction changes the dispersion of such a surface state by lifting the spin degeneracy, as shown in Fig. \ref{fig:3}(b). Formally, this can be described by the so-called Rashba model \cite{Rashba:1960,Bychkov:1984}. This model was originally developed to describe the interaction of a two-dimensional electron gas (2DEG) with an electric field perpendicular to the 2DEG's plane. A fast-moving electron in the plane will experience the Lorentz-transformed electric field as a magnetic field in the plane but perpendicular to its direction of motion. The electron's energy depends then on the orientation of the spin magnetic moment with respect to this magnetic field, parallel or anti-parallel, lifting the spin degeneracy in the dispersion. The model can be solved analytically, starting from from the free electron Schr\"odinger equation in two dimensions with the added energy for the spin-orbit interaction
\begin{equation}
H_{so}= -\frac{\hbar}{4 m_e c^2} (\mathbf{p} \times \mathbf{E}) \sigma =  -\frac{\hbar}{4 m_e c^2} (\nabla V \times \mathbf{p}) \sigma,
\end{equation}
where $\mathbf{p}$ is the momentum operator, $\mathbf{E}$ the electric field perpendicular to the surface with the generating potential $V$ and $\sigma$ the Pauli spin operator. The constants and the electric field strength can be represented by a parameter $\alpha$ and the interaction is added to the Hamiltonian of the free electron, leading to
\begin{equation}
 -\frac{\hbar^2 \nabla^2}{2m_e} \psi(\mathbf{r})+ \alpha (\mathbf{n} \times \mathbf{p}) \sigma\psi(\mathbf{r})= E \psi(\mathbf{r}),
\end{equation}
with $\mathbf{n}$ being a unit vector normal to the surface. This problem can be solved analytically and the resulting energies are 
\begin{equation}
E= \frac{\hbar^2 k^2}{2m_e} \pm \alpha \hbar k,
\label{equ:1}
\end{equation}
where the $\pm$ sign corresponds to the different spin directions. This  dispersion shown in Fig. \ref{fig:3}(b).

The analytically predicted dispersion for the Rashba effect has been confirmed by high-resolution ARPES measurements of the surface state on Au(111), the heaviest noble metal with the strongest spin-orbit interaction \cite{Lashell:1996,Reinert:2003}. If we, for now, just regard the ARPES intensity as a ``picture of the band structure'', the result of such a measurement as shown in Fig \ref{fig:3}(c) \cite{Nechaev:2009} fits excellently with the prediction of the model.

From equation(\ref{equ:1}) it can be seen that the state is always degenerate at the SBZ centre. This is not a coincidence or  a special feature of the Rashba model but a result of time-reversal symmetry which assures the so-called Kramers degeneracy. Consider a solution of the Schr\"odinger equation with wave vector and spin given by $(\mathbf{k}, \leftarrow)$. Time-reversal symmetry guarantees a degenerate solution with $(-\mathbf{k}, \rightarrow)$. For $\mathbf{k}=(0,0)$ the states must thus be degenerate. Time-reversal symmetry is assumed to hold at the surface but it could possibly be broken by a magnetic field.

\begin{figure}[b]
\includegraphics[scale=.65]{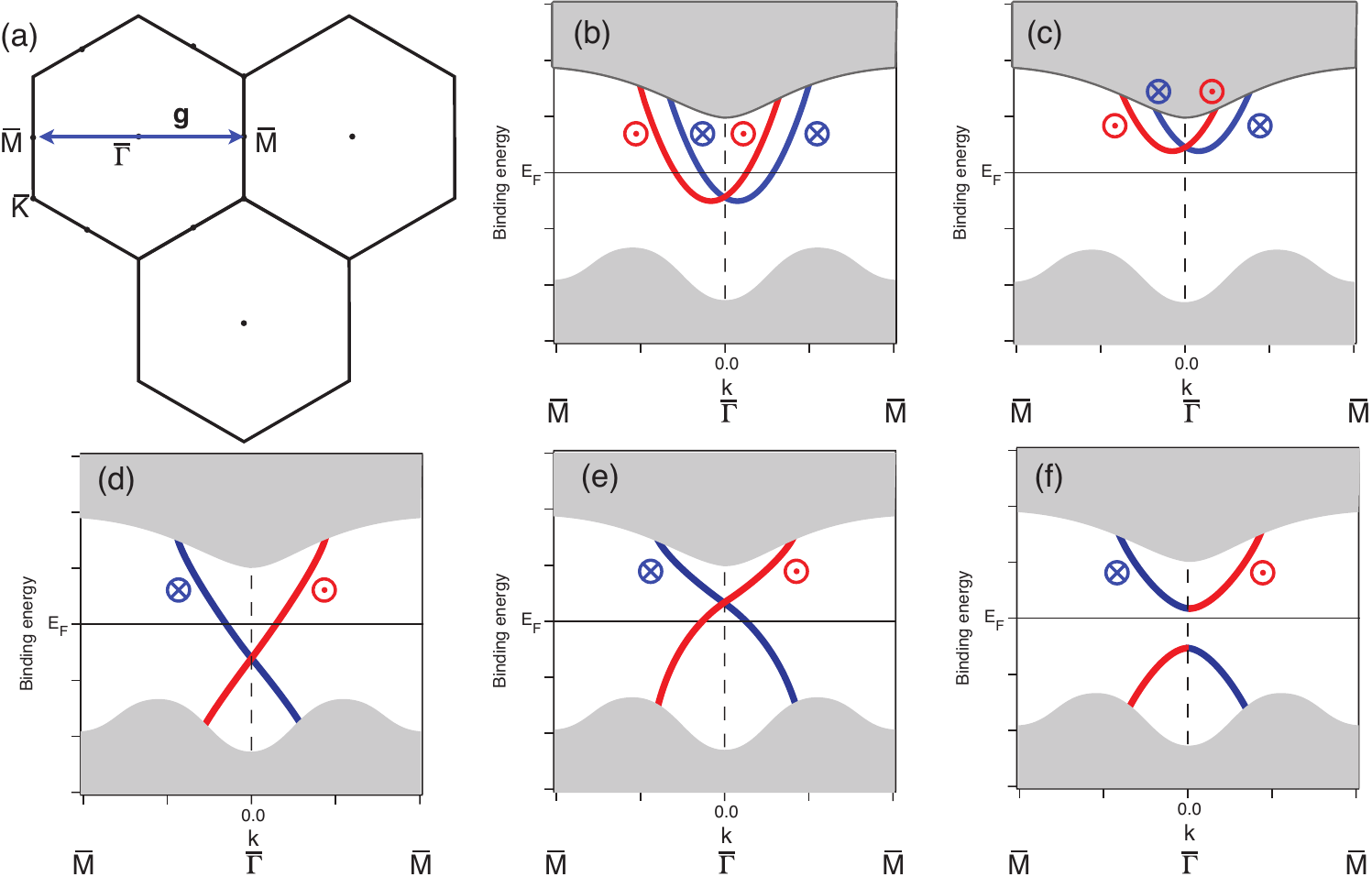}
%
%
\caption{(a) Hexagonal surface Brillouin zone with high-symmetry points. The $\bar{M}$ point can be identified as a time-reversal invariant momentum. (b,c) Rashba-split free electron-like surface state in a projected bulk band gap. The state in (b) is metallic but it can be emptied by lifting the entire dispersion above the Fermi level as in (c). (d,e) A topologically protected spin-split surface state. While the dispersion of the state depends on the details of the potential, time reversal symmetry protects it against a band gap opening, such that the situation in (f) cannot be realized. }
\label{fig:4}       
\end{figure}

Degeneracies in otherwise spin-split surface states can also be enforced at other points in $\mathbf{k}$-space, due to the combination of time-reversal symmetry and crystal symmetry. Consider for example a hexagonal SBZ in Fig. \ref{fig:4}(a) and a surface state with $(\mathbf{k}=\bar{M}, \leftarrow)$  at the $\bar{M}$ point. Time-reversal symmetry dictates a degenerate state with $(-\mathbf{k}, \rightarrow)$ but due to the symmetry of the lattice, $-\mathbf{k}$ is also an $\bar{M}$ point, equivalent to the staring point. Thus, there must be two degenerate states at this point (and every equivalent $\bar{M}$ point) with both spins, $\rightarrow$ and $\leftarrow$.

Points such as $\bar{M}$ in the hexagonal lattice are called surface time-reversal invariant momenta  ${\Lambda}_a$ (TRIM). They are characterized by the property  that 
${\Lambda}_a= -{\Lambda}_a+\mathbf{g}$ 
where $\mathbf{g}$ is a surface reciprocal lattice vector. There are always four possible TRIMS for the two-dimensional reciprocal lattice spanned by the  vectors $\mathbf{b}_1$ and $\mathbf{b}_2$. The reciprocal lattice vectors are given by $\mathbf{g} = n \mathbf{b}_1 + m \mathbf{b}_2$. One TRIM corresponds to $(m,n)=(0,0)$ and the other three are placed half-way to the points for which $(m,n)=(1,0),(0,1),(1,1)$, i.e. the remaining three independent combinations of the indices. 

We now discuss the topological stability from the perspective of the surface states in the presence of strong spin orbit splitting. Consider an insulator with a hexagonal Brillouin zone and the projection of the bulk bands along one direction, for example  $\bar{\Gamma}-\bar{M}$. This projection is shown as grey areas in Fig. \ref{fig:4}. The absolute band gap in a semiconductor is of course also reflected in a projected band gap around the  Fermi energy $E_F$. In Fig. \ref{fig:4}(b) we imagine that the surface hosts a free electron-like surface state, split by the Rashba effect. The surface state is only partially filled, crosses $E_F$, and the surface is therefore metallic. There is, however, no special topological protection of this metallic surface state. We could imagine to hole-dope the surface such that the entire dispersion is lifted above $E_F$, rendering the surface semiconducting as in Fig. \ref{fig:4}(c).

The situation is different for the surface state dispersion shown in Fig. \ref{fig:4}(d). In this case, a change in the dispersion could still be achieved by changing the potential near the surface. It would, for example, be possible to move the crossing point of the two spin-polarized branches above the Fermi energy as in Fig. \ref{fig:4}(e), but it would not be possible to open a gap in the dispersion as in Fig. \ref{fig:4}(f). This would violate time reversal symmetry that guarantees a spin-degenerate state at $\bar{\Gamma}$. The state in Fig. \ref{fig:4}(d) and (e) is thus a time-reversal symmetry protected state. Such states are found on the surfaces of the TIs and the dispersion of the states is very similar to the one shown here.

What is now special about the surface state in Fig. \ref{fig:4}(d) compared to the state in Fig. \ref{fig:4}(b)? It is not the dispersion as such. In fact, the dispersion of the state in Fig. \ref{fig:4}(d) could be a magnified version of the usual Rashba dispersion in the immediate vicinity of the crossing point, with the states disappearing in the projected bulk state continuum before the $k^2$ term in the dispersion of (\ref{equ:1}) becomes significant. The important difference is rather the number of Fermi level crossings between the SBZ centre and the SBZ boundary or, more precisely, between the surface TRIM at $\bar{\Gamma}$ and the surface TRIM $\bar{M}$ at the SBZ boundary. If the state shows an even number of Fermi level crossings between two surface TRIMS, it can be removed, but if it shows an odd number of crossings, it is topologically protected.

The important point is now that the number (even or odd) of Fermi level crossings between two surface TRIMs can be predicted solely by the bulk band structure of the material. A mere knowledge of the bulk bands thus permits statements about the existence of topologically protected surface states, not merely interface states as in the simple examples given in the beginning of this section. A second material is not required for stable states or, in a sense, vacuum can be viewed as a material with ``normal'' band ordering. Note that the topological considerations do not give a detailed prediction of the surface band structure, they merely predict if an odd or even number of surface state Fermi level crossings is present (actually, the predictions are somewhat stronger than this \cite{Teo:2008}, but this is of no concern here). Again the analogy with the bridge in Fig. \ref{fig:2} is useful. If we were told that a bridge had been installed that takes care of the traffic-flip problem, we could predict that there must be an odd number of crossings built into such a bridge but not how many (given common sense and budget restrictions, the answer would likely to be one).

\begin{figure}[b]
\includegraphics[scale=.45]{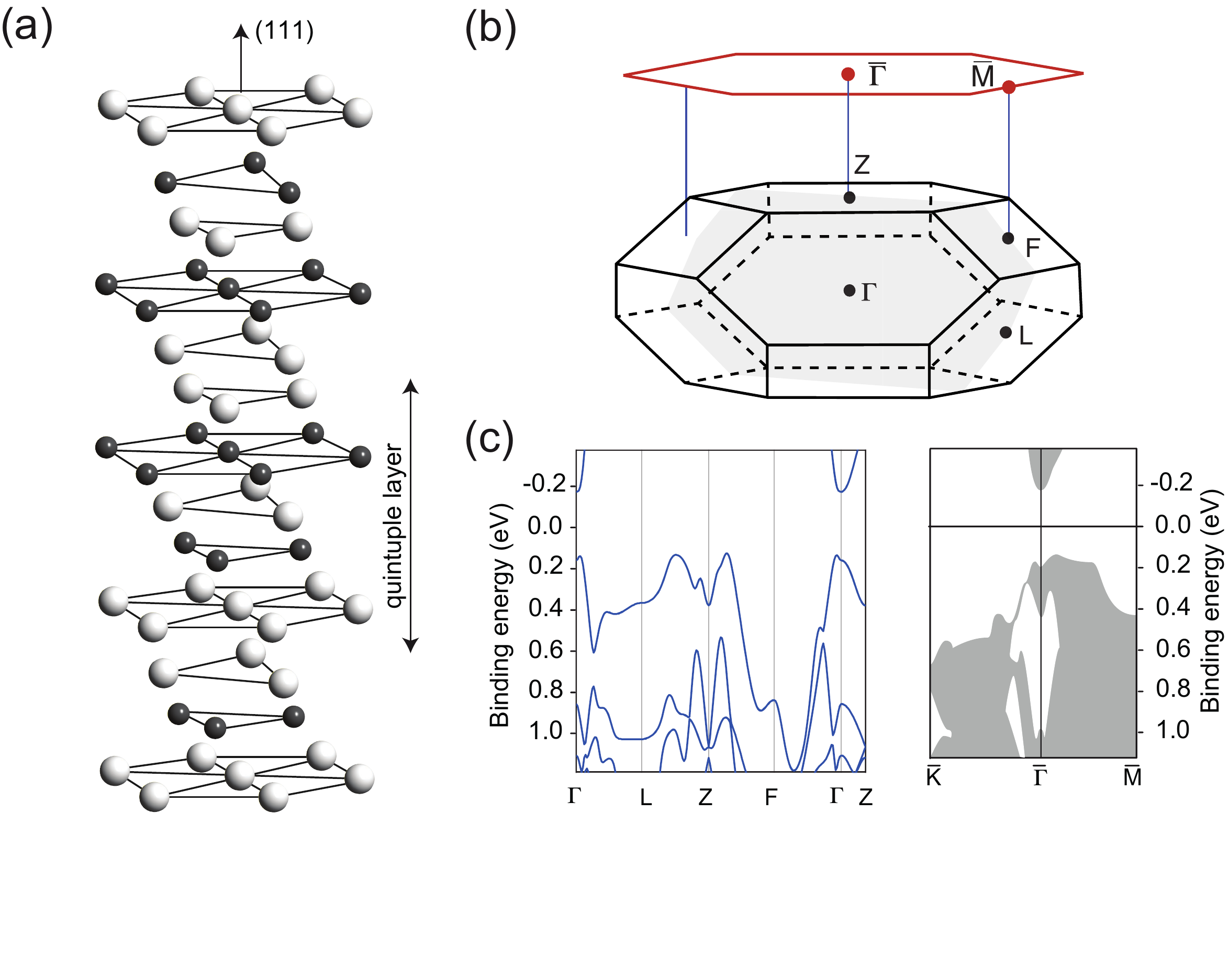}
\caption{(a) Crystal structure of Bi$_2$Se$_3$ with the quintuple layer building blocks. (b) Bulk and surface Brillouin zones with bulk time-reversal invariant momenta (TRIMs) and their projection to surface TRIMs. (c) Bulk band structure along selected high symmetry points and projection on the (111) surface after Ref. \cite{Eremeev:2010b}.}
\label{fig:4a}       
\end{figure}

We now illustrate how the number of surface state crossings between two surface TRIMs can be derived from the bulk band structure. 
This has been described in detail using the example of the TI Bi$_{1-x}$Sb$_{x}$ by Teo, Fu and Kane \cite{Teo:2008}. We follow their treatment here but we use the TI Bi$_2$Se$_3$ as an example. The bulk structure, bulk band structure as well as the bulk and surface BZ for the (111) surface of this material are given in Fig. \ref{fig:4a}.  

We first define the bulk TRIMs ${\Gamma}_i$ in analogy to the surface TRIMS by 
 $-{\Gamma}_i={\Gamma}_i+\mathbf{G} $ where $\mathbf{G} $ is a bulk reciprocal lattice vector. There are eight bulk TRIMS. In Bi$_2$Se$_3$ these are $\Gamma$, $Z$, three $F$ points and three $L$ points (see Fig. \ref{fig:4a}(b); note that the BZ actually contains two $Z$ points and six $F$ and $L$ points, but they are shared between two neighbouring zones). For each TRIM, the so-called parity invariants $\delta_i$ for the occupied bands are calculated by
\begin{equation}
	\delta \left ( {\Gamma}_i \right ) = \prod_{n=1} \xi_{2n}\left ({\Gamma}_i\right )
\label{equ:2}
\end{equation}
 where the $\xi_{2n}({\Gamma}_i) = \pm 1$ are the parity eigenvalues of the $2n$th occupied band at ${\Gamma}_i$, obtained from a bulk band structure calculation. Note that the bulk inversion symmetry of Bi$_2$Se$_3$ causes the bulk bands to be spin-degenerate because a state at $(\mathbf{k},\leftarrow)$ has a degenerate partner with $(-\mathbf{k},\leftarrow)$ due to inversion but also one with $(-\mathbf{k},\rightarrow)$ due to time-reversal. The index $n$ in equation (\ref{equ:2})  counts all the states, i.e. each spin-degenerate band is counted twice. The product is therefore only over every other (spin-resolved) band, such that every parity eigenvalue only appears once.
 
The topological character character of the bulk insulator is given by the so-called $\mathbf{Z}_2$ invariant $\nu_0$. For $\nu_0=1(-1)$ the material is a topological (trivial) insulator. $\nu_0$ can be calculated from the parity invariants  at the eight TRIMs by
\begin{equation}\label{nu0}
(-1)^{\nu_0}= \prod_{n=1}^8 \delta \left ( {\Gamma}_i \right ).
\label{equn:3}
\end{equation}

For Bi$_2$Se$_3$, we have 28 valence electrons per unit cell ($2\times5$ from Bi and $3\times6$ from Se), giving rise to 14 filled bands. The bulk parity invariants for the $Z$, $L$ and $F$ TRIMs are all calculated to be 1 but for the bulk $\Gamma$ point $\delta_i$ is found to be -1 \cite{Zhang:2009,Zhang:2010}. The product of (\ref{equn:3}) is thus found to be $-1$, hence $\nu_0=1$ and Bi$_2$Se$_3$ is established to be a topological insulator. 

The bulk parity invariants can also be used to describe fundamental properties of the surface electronic structure. To do this, 
 the so-called surface fermion parity $\pi_i$ can be determined for each surface TRIM. Essentially, $\pi_i$ is  obtained by projecting out the bulk parity invariants onto the corresponding surface TRIMs, using the relation $\pi(\lambda_a)=\delta(\Gamma_{i})\delta(\Gamma_{j})$. \footnote{Note that this treatment is somewhat simplified as the number of occupied bulk bands and the position of the surface cleavage plane can give rise to sign changes in the $\pi$ values. This affects the predictions about the detailed surface electronic structure but not the existence of the topologically protected states.}   We apply this to the (111) surface of Bi$_2$Se$_3$ with the SBZ shown in Fig. \ref{fig:4a}(b). In this case, $\pi(\bar{\Gamma})$ has to be calculated from the parity invariants of the bulk $\Gamma$ and $Z$ points, which are -1 and 1, respectively, and hence  $\pi(\bar{\Gamma})=-1$. For $\bar{M}$, on the other hand, $\pi(\bar{M})=\delta(L)\delta(F)=1$. Consequently, the surface fermion parity changes from $-1$ to $1$ along the path from $\bar{\Gamma}$ to $\bar{M}$. Such a change can be shown to imply an odd number of surface state Fermi level crossings along the line connecting the two TRIMS, and a closed Fermi contour around the surface TRIM with $\pi=-1$. These requirements are fulfilled by the electronic structure shown in Fig. \ref{fig:4}(d) and this electronic structure is, indeed, also observed for Bi$_2$Se$_3$.

\section{Angle-resolved photoemission spectroscopy (ARPES)}

ARPES has without question been a key-technique in identifying the topological surface states on TI and we will illustrate the power of the technique in the next section. Here we briefly discuss the principle of ARPES as far as required for the rest of this Chapter. For more detailed information about this well-established technique, the reader is referred to a number of excellent reviews and books \cite{Himpsel:1985,Plummer:1982,Kevan:1992,Matzdorf:1998,Hufner:2003,Hofmann:2009b}).

\begin{figure}[b]
\sidecaption
\includegraphics[scale=.55]{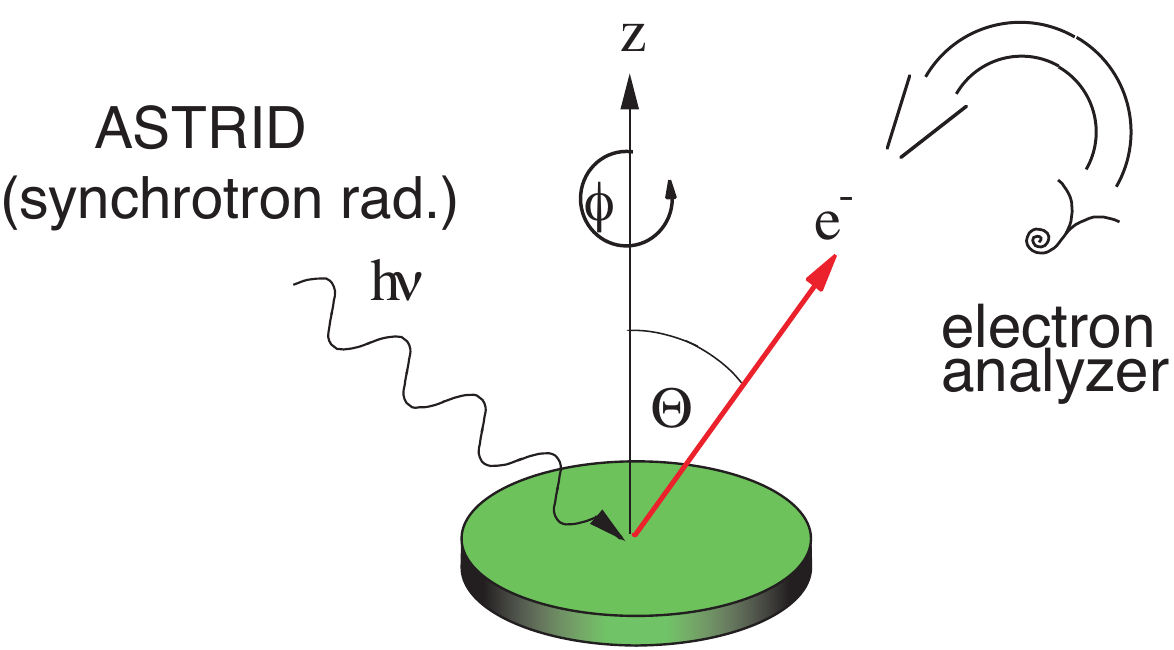}
%
%
\caption{Working principle of angle-resolved photoemission spectroscopy (ARPES). The technique is based on the photoelectric effect where the absorption of UV-photons leads to electron emission. The electron emission current is measured as a function of direction with respect to the sample surface normal  and as a function of kinetic energy. This is achieved by a hemispherical electron analyser with an entrance lens and an electron counter. It is also possible to measure the spin of the photoelectrons by replacing the electron counter with a spin-sensitive detector.}
\label{fig:5}       
\end{figure}

The working principle of ARPES is illustrated in Fig. \ref{fig:5}. Incoming UV photons cause the emission of photoelectrons from the solid and these electrons are detected by a spectrometer. The emission is only studied in a small range of solid angle, defined by the emission angles $\theta$ and $\phi$. The photoemission intensity of the emitted electrons is measured as a function of kinetic energy and intensity maxima in this distribution are assigned to emission from particular states in the sample. Thus, one measures the kinetic energy and the $\mathbf{k}$-vector for such states outside the surface and the objective of the analysis is to work back to the binding energy and $\mathbf{k}$-vector inside the solid, i.e. to the band structure. It is also possible to measure the spin of the photoelectrons and to use this in order to draw conclusions about the spin of the states in the sample. This technique of spin-polarized ARPES suffers from very inefficient detectors and therefore poor resolution and statistics. For identifying the spin texture of the topological surface states, however, it is indispensable.
 
It is useful to first establish the purely kinematic conditions for the observation of a state in ARPES. Energy conservation demands that emission from a state with binding energy $E_b$ (with respect to the Fermi level) leads to a photoelectron with kinetic energy $E_{kin} = h\nu - E_b - \Phi$, where $h\nu$ is the photon energy and $\Phi$ the sample's work function. Momentum conservation, on the other hand, is more involved because the  introduction of the surface breaks the translational periodicity of the crystal in the $z$ direction. The wave vector in that direction $k_z$ is thus no longer well-defined. The 
components of the wave vector parallel to the surface ($\mathbf{k}_\|$), on the other hand, are still well defined and must be  conserved in the photoemission process. $\mathbf{k}_\|$ of the electron outside the surface (and thus also from the state inside the surface) is obtained from
 \begin{equation}
\mathbf{k}_\|= (\sin{(\phi)} \mathbf{\hat{x}}+\cos{(\phi)}  \mathbf{\hat{y}}) \sin(\theta)\cdot \sqrt{\frac{2 m_{e} E_{{kin}}}{\hbar^2}}.
\end{equation}
For surface-localized states, such as the topological state, the dispersion of the band only depends on $\mathbf{k}_\|$; $k_z$ is not a relevant quantum number and the kinematic conditions are sufficient to extract the surface state dispersion from the measured photoemission intensity. 

\begin{figure}[b]
\sidecaption
\includegraphics[scale=.55]{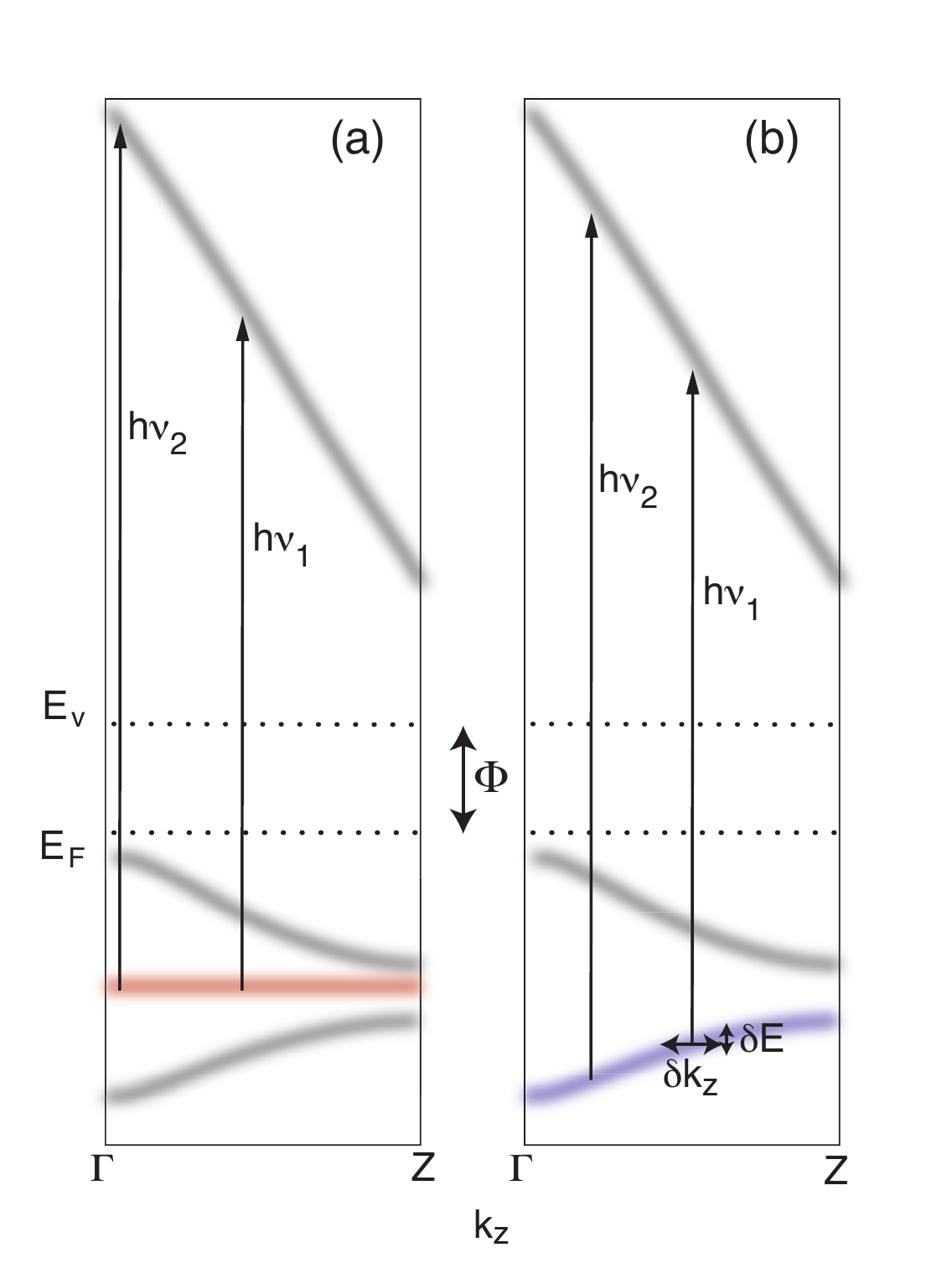}
%
%
\caption{Schematic picture of the photoemission process from different types of electronic states. $E_F$ and $E_V$ are the Fermi and vacuum level, respectively. $\Phi$ is the work function. Shown is the dispersion of different states as a function of wave vector perpendicular to the surface $k_z$. (a) The dispersion of a surface state does not depend on $k_z$ and therefore the observed binding energy for this state (distance from $E_F$) does not depend on the photon energy $h\nu$. (b)
Bulk states measured with different photon energies  appear at different binding energies in the spectrum, i.e. they show dispersion. The fact that $k_z$ is not well-defined leads to a broadening $\delta k_z$ that is then reflected in an energy broadening $\delta E$ of the observed peaks.}
\label{fig:7}       
\end{figure}

The fact that the surface state binding energy is independent of $k_z$ also implies that the state will be observed to be at the same binding energy, regardless of the photon energy used in the experiment. To see this, consider Fig. \ref{fig:7}(a) that shows the dispersion of a bulk initial state, a surface state, and a final state as a function of $k_z$. The surface state energy does not depend on $k_z$ but the energy of the two other states does. In the photoemission process, occupied initial states are excited into unoccupied final states and the photon energy thus determines the permitted $k_z$ for the transition from initial state to final state. For the surface state, the kinetic energy of the photoemitted electron depends on the photon energy but the measured binding energy (with respect to the Fermi energy) does not. The absence of dispersion upon a change in photon energy is thus a necessary condition for identifying a state as a surface state. 

As pointed out above, the wave vector in the $z$ direction is not a good quantum number near the surface because the translational symmetry in this direction is broken. Even if we ignore this fundamental problem for the time being, it would not be possible to recover $k_z$ inside the sample from the value measured outside because in order to do this the dispersion of the final states would have to be known \cite{Hufner:2003}. This is evident from Fig. \ref{fig:7}(b) that shows illustrates photoemission from a bulk band at two different photon energies. The $k_z$ value of the emitted electron depends on the dispersion of initial state and final state, as these define the $k_z$ value for which the photon energy corresponds to the energy difference between these bands. 

Despite these difficulties, it is often still possible to recover $k_z$ of the initial state from photoemission spectra taken as a function of $h\nu$, at least for high symmetry points in the BZ where the dispersion reaches an extremum. Once the photon energy for these extrema is known, a frequently used approach to recover $k_z$ for the remaining dispersion is the assumption of  free electron final states. $k_z$ plus or minus a reciprocal lattice vector can then be calculated  by
\begin{eqnarray}
k_z=\sqrt{{2m_e} /{\hbar^2} (V_0+E_{{kin}})},
\label{eq:kz}
\end{eqnarray}
where $V_0$ is the so-called inner potential.
 $V_0$ can be determined iteratively by requiring the resulting $k_z$ to be consistent with the binding energy extrema at high symmetry points. 

The problem that $k_z$ is not well-defined anymore cannot be neglected either. In particular, the finite escape depth of the photoelectron leads to a $k_z$ broadening $\delta k_z$. As the bulk state disperses in the $k_z$ direction, this also leads to an energy broadening $\delta E$, as illustrated in Fig. \ref{fig:7}(b). 

For surface states, and two-dimensional states in general, there is no additional broadening and this permits a much more far-reaching interpretation of the photoemission intensity. Indeed, the photoemission intensity can be can be interpreted in terms of the state's hole spectral function $\mathcal{A}$ weighted by the Fermi-Dirac distribution $f$ and a matrix element $|M_{fi}|$
  \begin{equation}
I(E_{kin},\mathbf{k}) \propto |M_{fi}|^2 f(h\nu-E_{kin}-\Phi,T)
\mathcal{A} (h\nu-E_{kin}-\Phi,\mathbf{k}),
 \label{equ:p2}
 \end{equation}
at least when infinitely good energy and angular resolution are assumed. The spectral function can be written as
 \begin{equation}
 \mathcal{A}(\omega,\mathbf{k},T)=\frac{\pi^{-1}|\Sigma''(\omega,\mathbf{k},T)|}{[\hbar \omega-\epsilon(\mathbf{k})-\Sigma'(\omega,\mathbf{k},T)]^{2}+\Sigma''(\omega,\mathbf{k},T)^{2}},
 \label{equ:A}
  \end{equation}
 where $\epsilon(\vec{k})$ is the single-particle dispersion.
$\Sigma'$ and $\Sigma''$ are the real and imaginary part of the so-called self-energy that contains the information about many-body effects. As evident from (\ref{equ:A}), $\Sigma'$ leads to a deviation of the state's dispersion from the single-particle case and $\Sigma''$ gives rise to a broadening that corresponds to a finite hole lifetime $\tau=\hbar / 2 \Sigma''$. The possibility to access the self-energy by ARPES thus gives direct access to the state lifetime, allowing the evaluation of many-body effects such as electron-defect scattering or electron-phonon scattering.

\section{Measured electronic structure of topological insulators}
\label{sec:3}

In this final section we show how the surface electronic structure of the prototypical TI Bi$_2$Se$_3$ can be determined by ARPES. We will illustrate how the topological states are identified, how their spin texture is confirmed by spin-resolved ARPES and how information about many-body effects can be obtained, using the example of the electron-phonon coupling.

\subsection{Observation of the topological surface states}

We address the situation for the (111) surface of Bi$_2$Se$_3$ which is the only surface of this material  that could be prepared so far. The reason why investigations are restricted to the (111) surface derives from the bulk structure of Bi$_2$Se$_3$ that is a stack of covalently bonded quintuple layers, separated by gaps with largely van der Waals bonding (see Fig. \ref{fig:4a}(a)). The (111) surface of the material can very easily be prepared by cleaving the sample parallel to the quintuple layers. However, cleaves in other directions have not been achieved yet. 

Turning back to the topological considerations, we have found that  Bi$_2$Se$_3$(111) shows a change of  surface fermion parity values between $\bar{\Gamma}$ with $\pi=-1$ and $\bar{M}$ with $\pi=1$. As stated above, this implies an odd number of Fermi level crossings between these surface TRIMS. Actually, the topological predictions are even stronger than this, requiring that the surface TRIM with the negative surface fermion parity must be enclosed by an odd number of Fermi contours. 

\begin{figure}[b]
\sidecaption
\includegraphics[scale=.55]{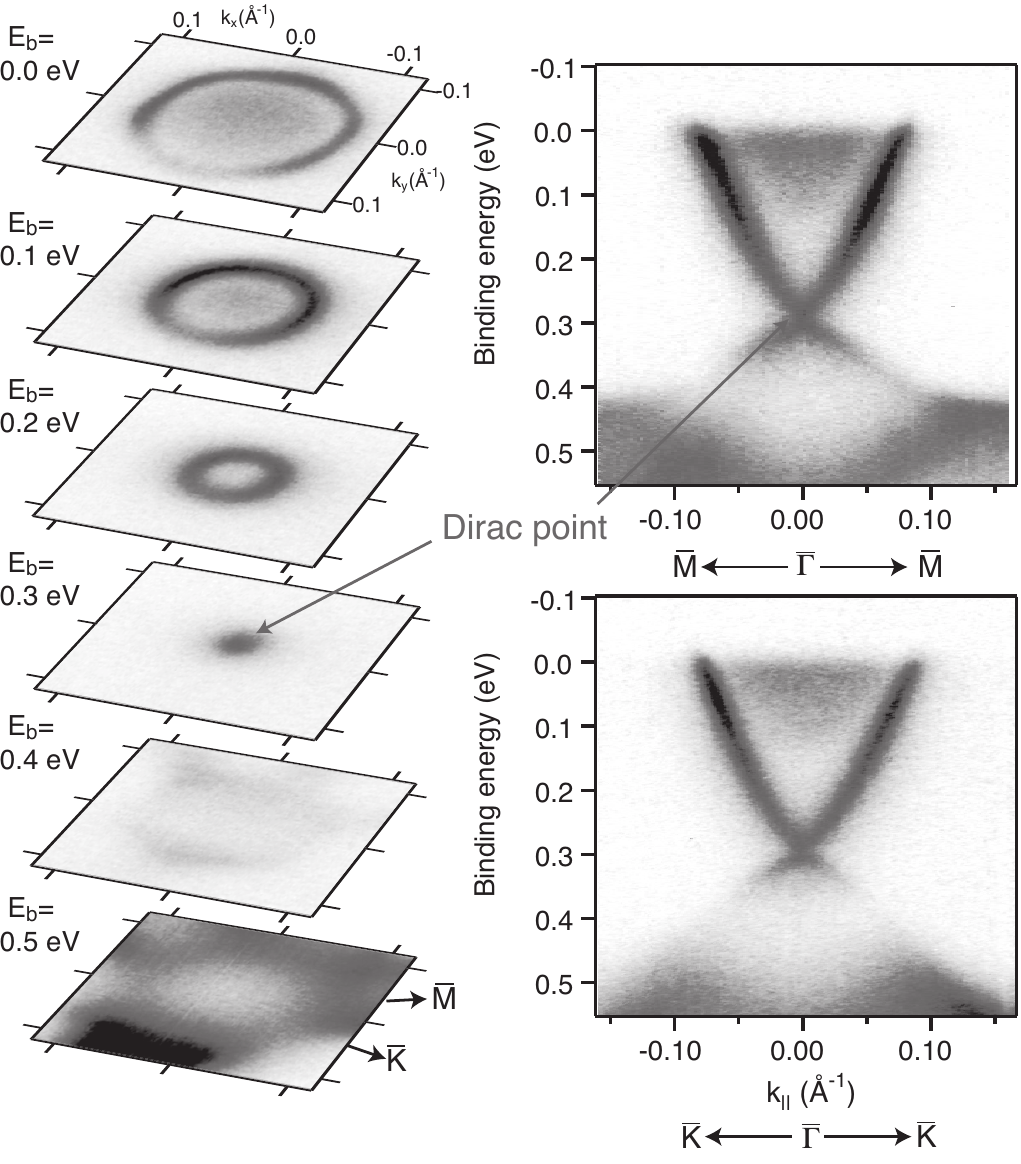}
\caption{Photoemission intensity from Bi$_2$Se$_3$ in the vicinity of the surface Brillouin zone centre. Different cuts through a three dimensional data set of the photoemission intensity $I(k_x,k_y,E_b)$ are shown. The left hand side shows cuts in $\mathbf{k}_\|$ at different binding energies. The right hand side shows the the dispersion of the states along two high-symmetry direction. The Dirac-cone shaped topological surface state can be easily identified, as well as states from the valence band and the conduction band (adapted from \cite{Bianchi:2010b}. }
\label{fig:8}       
\end{figure}

Fig. \ref{fig:8} shows cuts through a three-dimensional ARPES data set taken in the vicinity of the $\bar{\Gamma}$ point of Bi$_2$Se$_3$(111). Two types of cuts are show. On the left hand side the photoemission intensity is shown as a function of $\mathbf{k}_\| = (k_x,k_y)$ at fixed binding energies and on the right hand side it is shown as a function of binding energy and a specific high-symmetry direction in the $\mathbf{k}_\|$ plane. Clearly a cone-shaped state is identified that is very similar to the schematic state shown in Fig \ref{fig:4}(d). The state crosses the Fermi energy where the photoemission intensity drops to zero. At a binding energy of $\approx 0.3$~eV, the dispersion meets in the degeneracy point at $\bar{\Gamma}$. This particular dispersion and this point are often referred to as the Dirac cone and Dirac point, respectively. From such a data set alone, is very tempting to identify the Dirac cone as the topological surface state because of its metallic nature and its dispersion that is consistent with the topological predictions. 

Apart from the Dirac cone, broader features are observed at higher binding energies. These will be shown to derive from the uppermost valence band in Bi$_2$Se$_3$. Finally, a diffuse intensity at the Fermi energy is observed in the centre of the SBZ. This is caused by conduction band states, implying that this band is at least partly occupied. Indeed, the sample in question is degenerately $n$-doped, something that is frequently observed in pristine Bi$_2$Se$_3$ samples. This degenerate $n$-doping does not affect the possibility to observe the topological surface states by ARPES but it renders it very difficult to measure their contribution in transport experiments where the strongly doped bulk dominates. 

\begin{figure}[b]
\sidecaption
\includegraphics[scale=1.05]{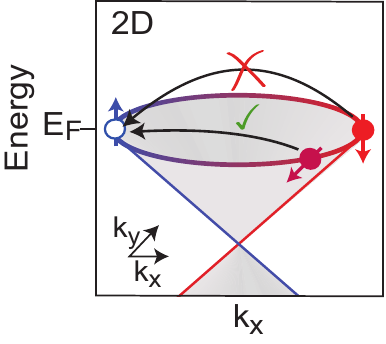}
\caption{Schematic representation of the surface state dispersion, the spin texture and possible scattering events for Bi$_2$Se$_3$. A hole at a constant energy surface on the left cannot be filled by elastic scattering of an electron at the opposite side of the Fermi contour because this process is spin forbidden. But it can be filled by scattering from any other state on the Fermi contour. }
\label{fig:9}       
\end{figure}

Before we discuss the further experimental evidence that the cone-shaped state is in fact the topological surface state, consider a sketch of this state in Fig. \ref{fig:9}(a). The state has a cone-line shape, it is not spin degenerate except for the $\bar{\Gamma}$ point and it is centred around this point. The general topology of this state is consistent the the predictions based on the surface fermion parity. There is an odd number (one) of Fermi level crossings between $\bar{\Gamma}$ and $\bar{M}$ and the $\bar{\Gamma}$ point is encircled by an odd number of Fermi contours (also one). No other surface states have been observed in the rest of the surface Brillouin zone of Bi$_2$Se$_3$ .

An important characteristic of the state is the non-degeneracy with respect to spin and the spin texture. The spin is expected to rotate on the constant energy surfaces of the cone while being perpendicular to the $\mathbf{k}_\|$ of the state. The sense of rotation is the same as for the inner branch of a Rashba-split state (see Fig. \ref{fig:4}), as the state can be viewed as derived from a Rashba state. Time-reversal symmetry for a non-degenerate state dictates that the spin of a state with $\mathbf{k}_\|$ is anti-parallel to that of the state with $-\mathbf{k}_\|$. Consequently, these two states are orthogonal and one expects a lack of backscattering in the system. In other words, a hole on one side of a constant energy surface cannot be filled by an electron on the opposite side (see Fig. \ref{fig:9}). This is a celebrated result for topological insulators and can be made visible by experiments with scanning tunnelling spectroscopy \cite{Roushan:2009,Zhang:2009a}. Not surprisingly, the same behaviour had also been found earlier for surface states on materials with very strong Rashba splitting in the surface states, which leads to an electronic structure that is very similar to that in the topological insulators \cite{Pascual:2004}. For the later discussion, it is important to note that, strictly spoken, only direct backscattering is spin-forbidden. Near-backscattering is merely unlikely (because the spin projection is still small) and near-forward scattering is hardly affected by the spin texture. Formally, this is described by a factor of $0.5 (1+\cos \alpha)$, where $\alpha$ is the angle between the two $\mathbf{k}_\|$ vectors involved in the scattering process \cite{Nechaev:2009}.

\begin{figure}[b]
\sidecaption
\includegraphics[scale=0.85]{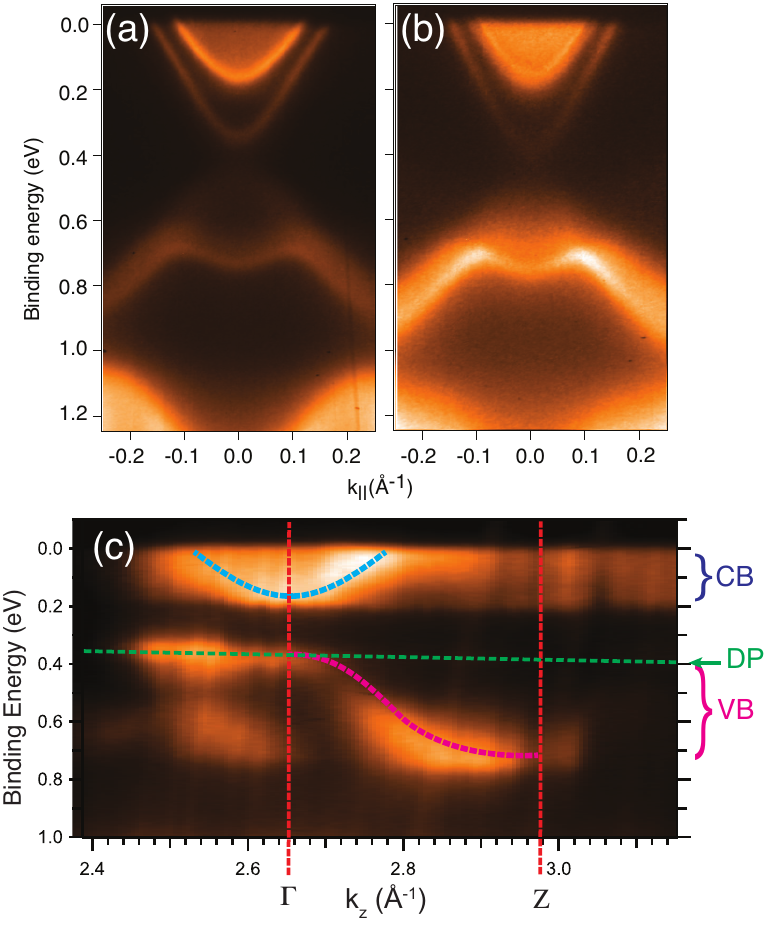}
\caption{(a,b) Photoemission intensity from the surface and bulk states in Bi$_2$Se$_3$ at two different photon energies  (19.2 and 26.6~eV, respectively). (c) Photoemission intensity in normal emission, showing the dispersion of the bulk CB and VB (emphasized by dashed lines). Note that the entire electronic structure gradually shifts to higher binding energies over time, an effect that is caused by contamination-induced band bending \cite{Hsieh:2009c,Bianchi:2010b}. The shift results in the small shift of the Dirac point (DP) (and the entire surface state dispersion) over the width of the scan. }
\label{fig:10}       
\end{figure}

The surface character of the observed state can be confirmed by performing experiments at different photon energies. According to the previous section, the surface state binding energy should not be affected by the choice of photon energy, in contrast to the energy of the bulk states. The result of such an energy scan is shown in Fig. \ref{fig:10}. The upper panel of the figure (a)-(b) shows the the photoemission intensity as a function of binding energy and $\mathbf{k}_\|$ for two photon energies (19.2 and 26.6~eV, respectively). Clearly, the topological surface state is observed at the same position, confirming the assignment as a two-dimensional state. The bulk VB and CB, on the other hand, change their appearance somewhat. This is more clearly seen in the data shown in Fig. \ref{fig:10}(c) that shows the photoemission intensity in normal emission only (i.e. the centre of the images in (a),(b)) as a function of binding energy and $k_z$. $k_z$ has been determined using free electron final states (equation (\ref{eq:kz})) with an inner potential of $V_0=11.8$~eV. In this representation of the data, the  dispersion of the CB and VB is clearly visible and high symmetry points can be identified ($\Gamma$ and $Z$). Such scans permit the determination of parameters such as the size of the band gap and width of the bands.

\begin{figure}[b]
\sidecaption
\includegraphics[scale=0.55]{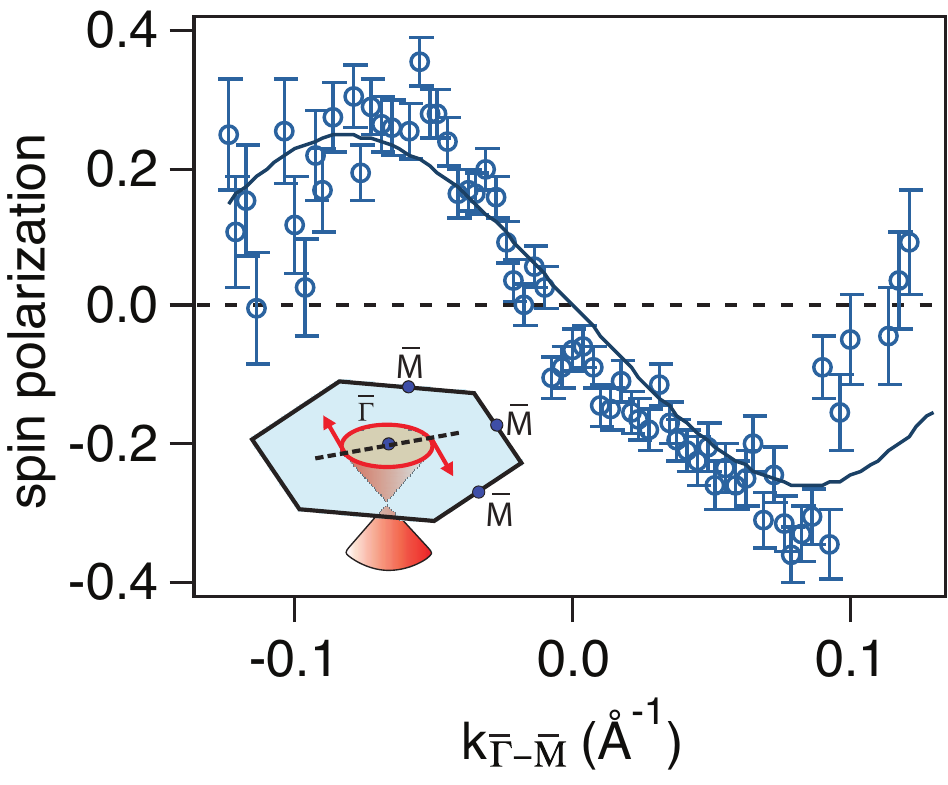}
\caption{Spin polarization of the topological surface state on Bi$_2$Se$_3$ measured by spin-resolved ARPES. The data show the determined polarization in the direction parallel to the surface and perpendicular to  $\bar{\Gamma}-\bar{M}$ at a binding energy $\approx$~100~meV above the Dirac point. The direction of the scan and the resulting spin polarization are indicated in the inset \cite{gabriel:tbp}. }
\label{fig:10a}       
\end{figure}

While the surface assignment of the topological state has thus been confirmed, we are lacking the confirmation of the expected spin texture and in particular of the non-degenerate character of the state. This is crucial for being able to assign the observed dispersion to a topological surface state rather than to an ordinary surface state. Fig. \ref{fig:10a} shows the result from a spin-resolved ARPES experiment  \cite{Hoesch:2002,gabriel:tbp}. The data points show the spin polarization in a scan along the $\bar{\Gamma}-\bar{M}$ direction that includes the two Fermi level crossings of the topological state, as indicated in the inset of the figure.  The degree of spin polarization is measured in the direction parallel to the surface but perpendicular to the scan direction. The states crossing the Fermi level are found to be strongly spin-polarized  but in opposite directions, consistent with the expected polarization in a Rashba model. This does not only confirm the expected spin texture but also the assumption that the state is not spin-degenerate, apart from at the $\bar{\Gamma}$ point. 

With this, it is firmly established that the observed state is in fact the predicted topological surface state. The combination of high-resolution spin-integrated ARPES and spin-resolved ARPES has been used to identify topological states on many different TI materials \cite{Hsieh:2009,Hsieh:2009b,Hsieh:2009c,Xu:2011}. 

\begin{figure}[b]
\sidecaption
\includegraphics[scale=.30]{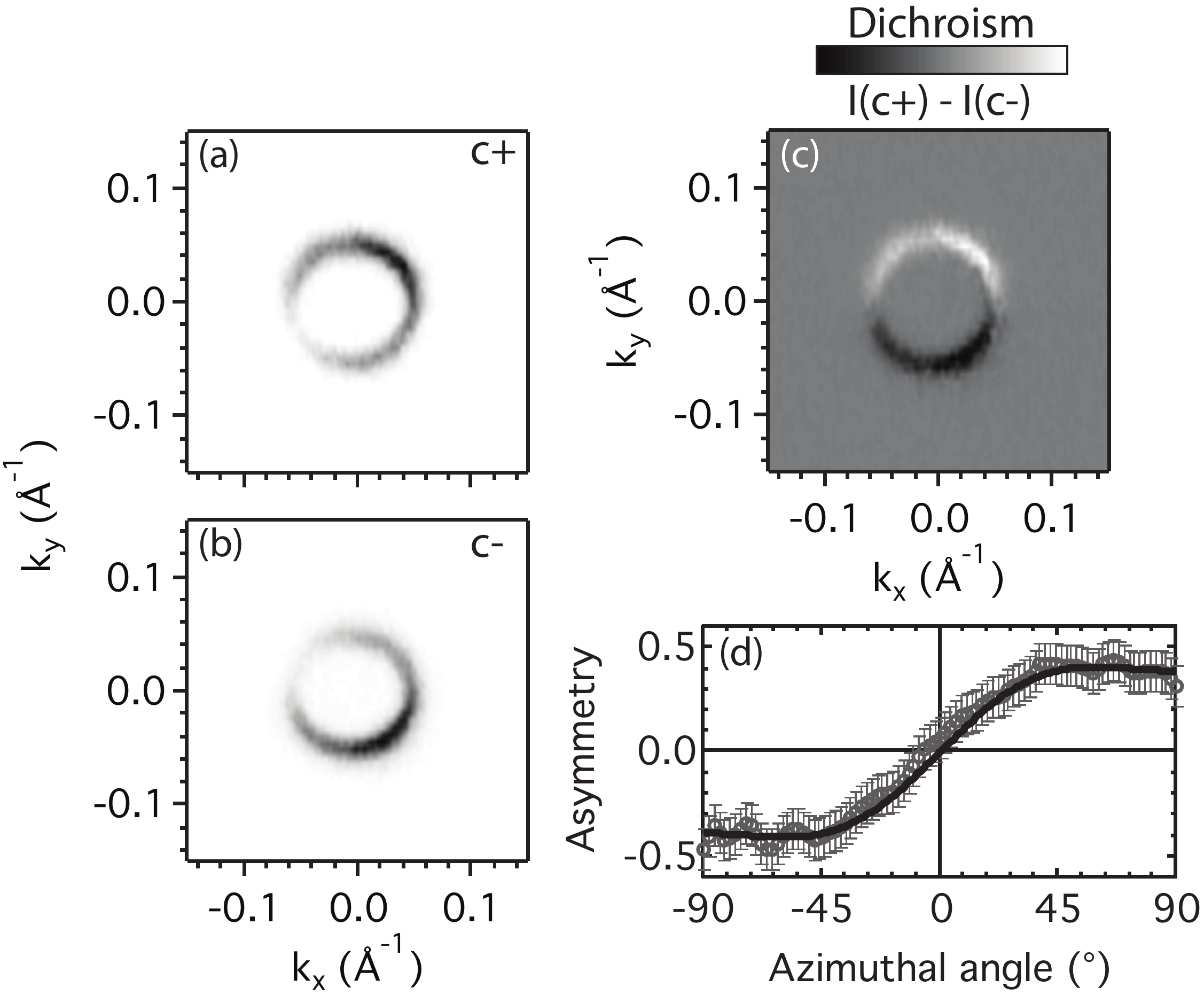}
\caption{Circular dichroism in ARPES from the topological surface state on Bi$_2$Se$_3$, approximately 150~meV above the Dirac point. (a,b) Photoemission intensity for light with right and left circular polarization, respectively. (c) Normalized difference between the two. (d) Quantitative dichroism along the circular contour. Data from \cite{Bahramy:2012}. }
\label{fig:13}       
\end{figure}

Recently, a number of studies have reported the observation of a strong circular dichroism in the ARPES intensity from topological surface states \cite{Park:2012,Park:2012b,Wang:2011c,Ishida:2011,Bahramy:2012} and it has been discussed if and how this can be related to the spin texture of the state. An example of the observed circular dichroism is given in Fig. \ref{fig:13} that shows polarization-dependent measurements of the photoemission intensity as a function of $\mathbf{k}_\|$ at an energy 150~meV above the Dirac point in Bi$_2$Se$_3$  ((a) and (b)), as well as circular dichroism obtained from these two data sets in an equivalent image (c) and quantitatively along the circumference of the circular contour (d) \cite{Bahramy:2012}. The strong effect of the light's polarisation is evident . The observed circular dichroism has been linked to the spin texture of the surface state and the technique has been proposed as an effective method to determine this spin texture. The detailed mechanism for this is still disputed but it appears likely that the circular dichroism is actually a consequence of the orbital texture rather than the spin texture. However, since spin and orbital degrees of freedom are very strongly coupled in a TI surface state, it may be possible to exploit the circular dichroism as a viable and efficient way to gain information about the state's spin texture.  

\subsection{Dynamics of the surface states: Electron-phonon coupling}

With the existence of the topological surface states firmly established, ARPES can be used to study the dynamics of these states. Of particular interest in this context is the sensitivity of the surface state electrons to defect scattering or electron-phonon scattering. After all, these processes limit the lifetime of excited carriers and thereby the surface channel conductivity. This is especially important because the bulk TI materials are found to be rather conductive, as explained above, and the desired transport situation is to have the surface state conductance dominate over the bulk conductance.

Possible scattering process for the surface state electrons were summarized in Fig. \ref{fig:9}. These processes are related to elastic scattering, i.e. defect scattering. The situation for electron-phonon scattering is not very different. Instead of taking the required momentum from an impurity scattering event, it is provided by the emission or absorption of a phonon. The process is not strictly elastic because the phonon energy has to be taken into account. However, since this energy is usually very small, it is often sufficient to consider the so-called quasi-elastic approximation where it is entirely neglected \cite{Hellsing:2002}. In this case, the situation is very similar to Fig. \ref{fig:9}. In particular, the phase space restriction due to the spin texture is identical.

\begin{figure}[b]
\includegraphics[scale=1.1]{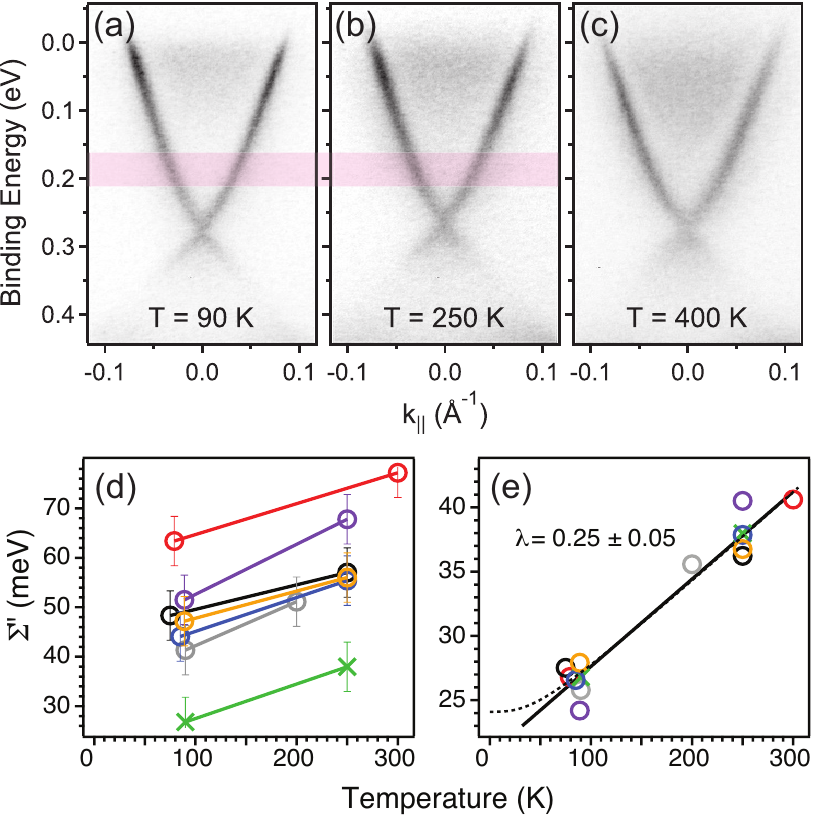}
\caption{Electron-phonon scattering on Bi$_2$Se$_3$ studied by ARPES. (a-c) Dispersion of the topological surface state at three temperatures. The shaded area indicates the energy range used for a quantitative determination of the linewidth. (d) Temperature-dependent width (expressed by the imaginary part of the self-energy) for different sample preparations. (e) The same data but with individual data sets rigidly shifted in order to account for different amounts of defect scattering. The dashed line shows a Debye model fit to the width and the solid line the high temperature (linear) limit. Data from Ref. \cite{Hatch:2011}.}
\label{fig:11}       
\end{figure}

In contrast to defect scattering, however, it is easy to probe the strength of the electron-phonon coupling by changing the number of available phonons for scattering processes via the sample temperature. Fig. \ref{fig:11}(a)-(c) show the dispersion of the surface state on Bi$_2$Se$_3$ measured at three different temperatures (after Ref. \cite{Hatch:2011}). The temperature effect over this range is not very big and the state appears sharp over the tested temperature range.

A more systematic analysis of the temperature-dependent width is given in Fig. \ref{fig:11}(d) that shows the temperature-dependent width of the state measured for many different sample preparations (via cleaving bulk crystals) but for only two temperatures per sample cleave. The reason why not more temperatures were measured is the rapid change of the surface electronic structure after cleaving. Two temperatures could be measured in a time interval short enough for this not to be an issue. 

The width of the state is expressed in terms of the imaginary part of the self-energy $\Sigma''$. This quantity is related to the inverse lifetime of the photohole via $\Sigma''=\hbar / 2 \tau$ and it is thus found that the lifetime increases at low temperatures, as expected. The data points have been collected in the energy range where the surface state is far away from the bulk bands, such that bulk-surface scattering cannot play a role. We see that the absolute $\Sigma''$ values for the different data sets are quite different but the difference between the two temperatures is similar. This is, in fact, to be expected because different sample cleaves lead to surfaces with a different amount of static defects. These give rise to an offset between the measurements but they do not affect the temperature dependence.

Fig. \ref{fig:11}(e) shows the same data but the different data sets have been shifted rigidly in energy in order to account for the different defect concentrations, and such that the data can be fitted by a theoretical model with a constant defect concentration. The theoretically expected $\Sigma''$ in the Debye model is shown as a dashed line and its high temperature limit as a solid line \cite{Hofmann:2009b}. Qualitatively, the shape of the dashed line is easy to understand: Far below the Debye temperature of the sample, the phonons are frozen out and the electron-phonon contribution to the (inverse) lifetime of the state is constant. At very high temperatures, on the other hand, the temperature dependence is linear, independent of the model used for the phonon dispersion. The slope is given by $2\pi k_B \lambda$, where $\lambda$ is the so-called electron-phonon mass enhancement parameter. The transition between these two regions is more complicated and depends on the model used. 

The result of this type of analysis is a value for $\lambda$, quantifying the electron-phonon coupling strength. In the analysis shown here, $\lambda=0.25(5)$ was found. To put this into context, $\lambda$ values for strong coupling BCS-type superconductors are found to be around $1$, whereas they are around $0.1$ for a weak-coupling good conductor, such as copper. 
It is thus evident that the coupling strength for the surface state of Bi$_2$Se$_3$ is not especially small. We can also compare the $\lambda$ value to a the fictitious situation of an isolated surface state, i.e. a surface state without the bulk present, such that only intra-state scattering is possible. This has been calculated for the well-studied noble metal surface states and for the Ag(111) state it has been found to be much weaker that the for Bi$_2$Se$_3$ ($\lambda=0.02$) \cite{Eiguren:2003}.

\section{Conclusion}

We have illustrated that synchrotron radiation-based ARPES is an essential experimental technique to study the surface electronic structure of TIs. It can be used to observe the topological surface state band dispersion and to distinguish these states from bulk states. Spin-resolved ARPES and possibly also circular dichroism ARPES can give essential information about the spin texture of the states. Further research on these materials will almost certainly involve other synchrotron radiation-based techniques that have not been discussed here. An interesting research field is the adsorption of potentially magnetic impurities on the surface, something that could locally break time-reversal symmetry. Here x-ray magnetic circular dichroism is an essential technique to determine the magnetic properties of the adsorbate \cite{Honolka:2012}. Another example is the structural determination of TI surfaces. The standard technique for this is low-energy electron diffraction but due to the complex structure and the distance of the interesting van der Waals gaps below the surface, it can be foreseen that synchrotron radiation-based surface x-ray diffraction may play an important role.

\section{Acknowledgement}
I gratefully acknowledge discussions with many students and colleagues who helped me to understand the physics of the topological surface states. I thank Justin Wells for making me aware of the ``flipper bridge'' in Fig. \ref{fig:2} and Kamiel Klaasse from NL Architects for letting me use this picture. I also like to thank Gabriel Landoldt and Hugo Dil for providing me with unpublished data for Fig. \ref{fig:10}, Phil King for Fig. \ref{fig:13} and Edward Perkins for a careful proofreading of the manuscript.


\end{document}